\documentclass[prl,floatfix,twocolumn,showpacs,amsmath,amssymb]{revtex4}
\usepackage{graphicx}% Include figure files
\usepackage{dcolumn}% Align table columns on decimal point
\usepackage{bm}% bold math

\newcommand{\ba}{\begin{eqnarray}}
\newcommand{\ea}{\end{eqnarray}}
\def\be{\begin{equation}}
\def\ee{\end{equation}}
\def\n{\nonumber\\}

\begin{document}

\title{Incidence of the boundary shape in the tunnelling exponent of
electrons into fractional quantum Hall edges}

\author{D.C.\ Cabra$^{1,2,3}$ and N.\ E.\ Grandi$^2$}

\affiliation{~~ \\
$^1$Laboratoire de Physique Th\'eorique, Universit\'e Louis
Pasteur, 3 rue de l'Universit\'e, F-67084 Strasbourg Cedex, France
~~\\
$^2$Departamento de F\'{\i}sica, Universidad Nacional de la Plata,
C.C.\ 67, (1900) La Plata, Argentina
~~\\
$^3$Facultad de Ingenier\'\i a,  Universidad Nacional de Lomas de
Zamora, Cno. de Cintura y Juan XXIII, (1832) Lomas de Zamora,
Argentina.}
\date{\today}

\date{\today}

\begin{abstract}

Experiments on tunneling into fractional quantum Hall droplets 
systematically found tunneling exponents smaller than those predicted 
by the ordinary chiral Luttinger liquid theory. In this note, by 
considering the effects of a smooth boundary, we propose a 
modification of the theory that predicts a reduced exponent, in 
qualitative agreement with experiments. Such modification consists 
on a self-interaction cubic term, containing higher derivatives of the chiral boson field, 
that provides a non-universal one loop correction to the propagator.

\end{abstract}

% \pacs{...}

\maketitle

The study of tunnelling of electrons into both incompressible and
compressible quantum Hall states has been the subject of intense
research from both the theoretical and experimental sides (see
\cite{horda} and references therein). It is found that the
tunnelling conductivity is non-Ohmic, $I \propto V^\alpha$, with
$\alpha$ being a function of the filling fraction $\nu$. This
behavior can be understood within the chiral Luttinger liquid
description of the edge physics advanced by Wen in \cite{Wen}.
Although tunnelling experiments in fractional quantum Hall effect
(FQHE) systems have shown certain degree of agreement with the
theoretical predictions obtained from the chiral Luttinger liquid
picture \cite{Wen}, there remain certain discrepancies which have
been addressed by different authors \cite{horda,first
diff,Yang,Wan,plateau,exp-num,Chamon-Fradkin}. In particular, a
reduction of the order of $10\%$ from the theoretical prediction
for the tunnelling exponent has been observed experimentally, as
well as the absence of the theoretically predicted plateaux
structure of this exponent \cite{first
diff,plateau,Chamon-Fradkin}. These issues have been the subject
of intense debate \cite{horda}.

Here we present an alternative derivation of the boundary
effective action by including higher order terms in the low energy
expansion which are induced by the smoothness of the boundary. The
effects of the edge shape and the confining potential on the
tunnelling exponents have been recently addressed using numerical
exact diagonalization \cite{Wan}, where it was found that
tunnelling exponents could deviate from the chiral Luttinger
liquid universal predictions. Our results, obtained analytically
starting from a microscopic description, are in complete agreement
with these findings.

More precisely, we start from the microscopic description of the
electron fluid in a strong magnetic field following
\cite{Susskind} and in order to include the effects of a smooth
boundary, we consider the space dependence of the electron density
over a boundary region of certain width $W$. This leads
generically to a boundary effective theory of $N$ coupled chiral
bosons, with $N = W/a$, $a$ being the mean interparticle distance.

In the simplest case of an almost sharp edge, the effective action
we obtain is that of a self-interacting chiral boson. Using this
improved effective action we compute the tunnelling exponent
which, to one loop order, receives a small non-universal negative
correction which depends on the electron density and the short
distance cutoff (the "size" of the particles). By suitably
choosing the UV cutoff, essentially as the mean distance between
particles, we show that our results are in agreement with the
experimentally observed values. Since our theory describes gapped
bulk excitations, the present approach is applicable within the
plateaux regions of the quantum Hall effect.

We first derive the boundary effective theory following
essentially the same steps as in \cite{Susskind} but with a smooth
boundary, the presence of which is shown to be crucial {\it
e.\,g.} in the computation of tunnelling exponents. Let us briefly
review the main steps in the derivation:

The action of a two dimensional system of charged particles
(charge $-e$) moving in a constant magnetic field $A_a =-\frac B
2\epsilon_{ab}x^b$ and an electric potential $V(x^a)$ is

\be
S =\sum_p\int dt\left( \frac m2 \, {\dot x^{a 2}_p} +\frac
{eB} 2 \,\epsilon_{ab}\,{\dot x}_p^a x^b_p-eV(x^a_p)\right)
\label{action1}
\ee
where $x_p^a$ denote the particle positions, $a=1, 2$ being the
space directions and $p$ being a discrete particle index.

The electric potential will be chosen to be approximately constant
inside some bounded region ${\cal M}$, vanishing smoothly outside
the boundary $\partial {\cal M}$ through a region ${\cal E}$ of
width $W$. This region, in which the electric field
$E_a=\partial_a V$ is non-vanishing, defines the edge of the
sample.
%and it is perpendicular to it.
The density $\rho_0$ will be chosen, consistently with the assumed
form of the sample, as approximately constant inside ${\cal M}$
and decreasing within the edge ${\cal E}$ to be zero outside.

To see the effect of $V(x^a)$ at the edge, we approximate it in
the neighborhood of each particle by a linear function
$V(x^a)=-E_a x^a$, assuming that the length in which changes in
the electric field are measurable is much larger than the cyclotron
radius, {\it i.e.} $r<<E/|\nabla E|$. Then we have an effectively constant
electric field
around the particle and we see
that it can be eliminated through the boost $\tilde
x^a= x^a+v^a t$, where $v_a = \epsilon_{ab} E_b/B$. Then the
classical
motion is given by circular orbits with cyclotron frequency
${eB}/m$ and radii related to the initial velocities by $r={|\dot
x_{in}| m}/{eB}$,
% and with
the centers
of which are
moving with velocity $-v^a$
transverse to the electric field, {\it i.e.} parallel to the edge.

As a conclusion, for a strong magnetic field the action can
be written as

\be S =
%\sum_p\frac {eB} 2 \int dt\,\epsilon_{ab}\,({\dot x}_p^a +v^a_p)(x^b_p + v^b_pt)=
\sum_p\frac {eB} 2 \int dt\,\epsilon_{ab}\,\dot {\tilde
x}_p^a\tilde x^b_p \ee
where the new variables $\tilde x^a_p$ are given by the above
expressions in terms of the position of each  particle and the
electric field in its neighborhood. Note that the velocity $v^a_p$
is parallel to $\partial{\cal M}$, being zero for the particles
outside ${\cal E}$.

If the mean distance between the particles $a=\sqrt\rho_0$ is
small compared to all the external scales involved into the
problem ({\em i.e.} the lengths at which variations on the
electric field or on the density are measurable), we can take the
continuum limit. Then, we can replace the variables $\tilde x_p^a$
by fields $\tilde x^a(\tilde y^i,t)$ where the continuous label
$\tilde y^i$ is fixed by the condition that $\tilde x^a(\tilde
y^i,t=0)=\tilde y^a$. Then the sum on $p$ can be replaced by an
integral

\be
S = \int  \!dt \,d^2\tilde y \,\rho_0\,\frac{eB}2  \,
\epsilon_{ab} \,\partial_t\tilde {x}^a \tilde x^b
\label{contaction}
\ee

In the continuum theory, the aforementioned condition can be
characterized by $E/\partial E>>a$ and $\rho_0/\partial
\rho_0>>a$, where $\partial$ represents the derivative in some
spatial direction. Since these derivatives can differ
significatively when taken in different directions, so does the
accuracy of the limit.

To be concrete: at the edge, if the curvature of the boundary is
small, then we have $E/\partial_{\parallel}E>>a$ (and the same for
$\rho_0$), and the limit works in the longitudinal direction. On
the other hand, for a sharp enough edge we may have
$E/\partial_{\perp}E\sim a$, and we most somehow pay attention to
the discrete structure on the transverse direction. We will
accomplish this bellow by re-discretizing the transverse
integration.

Here it is important to stress that $\tilde y$ is the initial
position of the fluid element in its particular ``rest frame'' in
which the electric potential term vanishes. Note that the operation
of taking the continuum limit introduces
the function $\rho_0$, which was part of the initial conditions,
into the action of the system in the form of a measure in space.

The resulting action is invariant under the group of measure
preserving diffeomorphisms in the $y^i$ plane, {\it i.e.} changes of
coordinates such that $y_i'= y_i +\epsilon_{ij}\partial_j \aleph /\rho_0$.
The corresponding conserved
generator is
\be Q = \frac{eB}2\int
d^2y\,\epsilon_{ab}\epsilon^{ij}\partial_j\tilde x^b\partial_i
\tilde x^a \aleph \ee
since it is conserved for any function $\aleph $, its integrand
is conserved, {\it i.e.}
\be
\epsilon_{ab}\epsilon^{ij}\partial_j\tilde x^b
\partial_i\tilde x^a=\left.\epsilon_{ab}\epsilon^{ij}\partial_j\tilde x^b
\partial_i\tilde x^a
\right|_{t=0}=2
\label{contr}
\ee
this relation is a constraint in the values of the fields $\tilde
x^a$ that has to be solved to obtain the independent dynamical degrees of
freedom of the system.

The next step is to solve the constraint (\ref{contr}) and rewrite
the action in terms of the dynamical degrees of freedom. Here and
in what follows, we will use as parameters the filling fraction
$\nu = \rho_0/eB$ and the effective non-commutativity parameter
$\theta = 1/(2\pi\rho_0)$ \cite{Susskind}.

If we parameterize the $\tilde x^a$ field with the help
of a new field $\phi$ according to
\be
\tilde x^a=y^a + \theta\,{\epsilon^{ab}}\partial_b\phi
+ \frac {\theta^2}2\,
\epsilon^{ab}\epsilon^{ij}\,\partial_i\phi\,\partial_j\partial_b\phi
+{\cal O}(\theta^3)
\label{sol}
\ee
then (\ref{contr}) is solved up to order $\theta^3$ in the
interior of ${\cal M}$. We see in (\ref{sol}) that $\phi$ is
defined up to the addition of an arbitrary function of time
$\phi\to\phi+g(t)$, and that its derivatives account for the
deviation of the particles from the classical center of their
motion.

At the edge, we take (\ref{sol}) as a parametrization of
$x^a$, which is a solution of the constraint up to subleading contributions containing
derivatives of $\theta$.

Replacing this solution in the action (\ref{contaction}) we obtain

\ba
S &=& \frac{1}{8\pi^2} \int  \!dt
d^2y\,\nu^{-1}\!{\epsilon^{ij}}\!\left(
\partial_{j}\phantom{{x^2}^2}\!\!\!\!\!\!\!\!\!\!
\left(\phi\,\partial_{i}\partial_t \phi\right) +
\right.
\n&&
\ \ \ \ \ \ +\,\frac \theta 3\left. \partial_{j}\!
\left(\epsilon^{b{c}}\partial_{b}
\partial_t \phi\,\partial_i\phi\,\partial_{c}\phi
\right) \phantom{{x^2}^2}\!\!\!\!\!\!\!\!\!\!\right)\,+\,{\cal O}(\theta^2)
\label{aa}
\ea

We reintroduce the effect of the electric field by boosting back
$y^i = \tilde y^i - v^it$, with the velocity defined above. We end
up with the Lagrangian description of a fluid, where each fluid
element is labelled with its initial position and it is followed
along its motion through the plane. Here the velocity is a
function of $y^i$, almost constant within ${\cal E}$ and vanishing
outside the sample.

In the special case in which $\nu, \theta$ are constants the
action is a total derivative. The usual procedure is to assume
this and then impose a boundary to the region in which the fluid
moves. We choose a different way here, allowing the parameters to
change in space and introducing through them the information about
the geometry, without adding any boundary. A further partial
integration gives

\ba
S &=& - \frac{1}{8\pi^2} \int  dt
d^2\!y\,{\epsilon^{ij}}\!\left(\phantom{{x^2}^2}\!\!\!\!\!\!\!\!\!
\partial_{j}(\nu^{-1})\,
\phi\,\partial_{i}(\partial_t-v^a\partial_a)\phi \,+
\right.
\n
&&\ \ \ \ +\frac{1}{3} \left.
\partial_{j}(\theta\nu^{-1})\,
\epsilon^{b{c}}\partial_{b}(\partial_t-v^a\partial_a)
\phi\,\partial_i\phi\,\partial_{c}\phi
\phantom{{x^2}^2}\!\!\!\!\!\!\!\!\!\!\right)
\label{aaa}
\ea
Here we see that, wherever the sample is homogeneous, the
derivatives in front of each term vanish, leaving us without any
dynamics. On the other hand, the dynamical degrees of freedom
localize at places where there is a change in the properties of
the material (noncommutativity and/or filling fraction). This is
precisely what happens at the edge of the sample, and it is at the
core of our argument

To stress this point, let us suppose that the parameters change as a
step function, then their derivatives will provide delta functions
leading us to a one dimensional boundary theory. But we see that our treatment
is more general in the sense  that it includes the possibility of
smooth edges.

The action (\ref{aaa}) is written completely in terms of the
dynamical degrees of freedom $\phi$. Note that this is not a
boundary theory because it is defined in the full two dimensional
space. Nevertheless, the degrees of freedom are bounded to the
region ${\cal E}$ in which there is a change in the parameters $\nu$ and
$\theta$ ({\it i.e.} in the density), and propagate chirally along their
level surfaces.

To proceed further, we define a strip in this setup by the
condition that the parameters are independent of one of the
cartesian coordinates $y^i = (x,y)$, say $x$. Then the action
becomes
\ba
S = -\frac{1}{8\pi^2} \int  \!dt \,dx dy\!\left(\phantom{{x^2}^2}\!\!\!\!\!\!\!\!\!
\partial_{y}(\nu^{-1})\,
\phi\,\partial_{x}(\partial_t-v\partial_x)\phi \,+
\right.
\n
+ \,\frac 1 2
\left. \partial_y(\theta\nu^{-1})\,
\partial_{y}\phi\,(\partial_t-v\partial_x)(\partial_{x}\phi)^2
\phantom{{x^2}^2}\!\!\!\!\!\!\!\!\!\right)
\label{ccc}
\ea

To recover the discrete structure in the direction $y$ transverse
to the edge,
% In order to perform the $y$ integration
we proceed
as follows: \footnote{This procedure may have some connection with
the so-called boundary reconstruction effect, which has been studied
from both the experimental and analytical point of views in
\cite{exp-num}.}

\begin{itemize}

\item Cut the space in the $y$ direction in slices at $y_n$ of
width $\Delta y \sim a$.

\item Replace the integral by a sum over the values of the
integrand evaluated at each slice, and define
$\phi_n(x)\equiv\phi(x,y_n)$.

\item Replace all the $y$ derivatives by its finite difference
approximation $\partial_y \phi(x,y_n) =
(\phi(x,y_n)-\phi(x,y_{n-1}))/\Delta y$.

\end{itemize}

With this method we obtain an effective theory for the fields
$\phi_n(x)$ living on each slice $n$, with action
\ba
S&=&-\frac{1}{8\pi^2}  \int  \!dt \,dx \sum_{n=0}^N
\left(\,\phantom{{x^2}^2}\!\!\!\!\!\!\!\!\!\!\kappa_n
{\cal L_F}(\phi_n) \,+
\right.
\n&&\ \ \ \ \ \ +
\left.
\kappa'_n \left({\cal L_S}(\phi_n) +
{\cal
L_I}(\phi_n,\phi_{n-1}) \phantom{f^F}\!\!\!\!\!\!\!\right)
\phantom{\frac12}\!\!\!\!\!\right)
\label{c}
\ea
where we have defined the free, self interaction and interaction
Lagrangians as
\ba
{\cal L_F}(\phi_n)&=&
\phi_n\,\partial_x(\partial_t-v_n\partial_x)\phi_n
\n
{\cal L_S}(\phi_n)
&=&
\phi_n(\partial_t-v_n\partial_x)(\partial_x\phi_n)^2
\n
{\cal L_I}(\phi_n,\phi_{n-1})
&=&
-\phi_{n-1}(\partial_t-v_n\partial_x)(\partial_x\phi_n)^2
\ea
and the constants $\kappa_n$ and $\kappa'_n$ are given by
\be
\kappa_n =\frac 1{\nu_n}-\frac 1{\nu_{n-1}} \ , \ \ \ \ \
\ \ \kappa'_n =\frac 1{2\Delta y}\left(\frac {\theta_n}{\nu_n}-
\frac{\theta_{n-1}}{\nu_{n-1}}\right) \ .
\label{levels}
\ee
It is important to note that the field $\phi_n$ will enter into
the action only when the constants $\kappa_n, \kappa'_n$ are
non-vanishing, {\it i.e.} if there is a change in the properties of the
sample between the slices $n$ and $n+1$.

The gauge invariance is now $\phi_n\to\phi_n+g(t)$ adding the same
$g(t)$ to all the $\phi$'s.

Note that when the sample has a sharp edge, {\it i.e.} if the density
is changing in a region whose width is smaller that the slicing length $W<\Delta y
\sim a$, then the whole procedure is not applicable and the boundary
theory corresponds to the usual chiral boson theory \cite{Wen}.

Let us suppose that the edge is wider than the slicing length, so
that we have to keep a finite number of terms of the sum in
(\ref{c}), $n=1,\cdots , N$.

We have then a single non-dynamical field $\psi\equiv \phi_{-1}$
in the interaction term ${\cal L_I}(\phi_0,\phi_{-1})$, all other
fields appearing in the ${\cal L_I}(\phi_n,\phi_{n-1})$ terms
being dynamical. The integration of this multiplier field will
enforce a constraint on the field $\phi_0$ to which it is coupled.
When solved, $\phi_0 = f(x+v_0t)$, and replaced in the action, it
implies that the interaction term with $\phi_1$ takes the form
\ba
 -\frac{1}{8\pi^2} \kappa'_1
\int  \!dt \,dx
f(x+v_0t)(\partial_t-v_1\partial_x)(\partial_x\phi_1)^2 \label{cc}
\ea
which is different from zero provided that the two velocities are
not equal. Since we have assumed that the electric field in the
edge zone is approximately constant in regions of size $a\sim\
\Delta y$, hence $v_1 \approx v_2$ and we can discard this term within the present approximation.

We have then obtained an action very similar to the original one but
without any Lagrange multiplier
\ba
S= -\frac{1}{8\pi^2} \int \!dt \,dx \sum_{n=1}^N\left(\phantom{{x^2}^2}\!\!\!\!\!\!\!\!\!
\kappa_n {\cal L_F}(\phi_n) + \kappa'_n \left({\cal L_S}(\phi_n) \,+
\right. \right.
\n+
\left. \left.
(\delta_{n1}-1){\cal L_I}(\phi_n,\phi_{n-1})
\phantom{2^2}\!\!\!\!\!\right)\phantom{\frac12}\!\!\!\!\!\!\right)
\label{cb}
\ea
where the lower bound of the sum has changed and the coefficient
$(\delta_{n1}-1)$ ensures that there is no interaction term for
the field $\phi_{1}$.

Let us consider the case of an almost sharp boundary, in which the
density can be well approximated by choosing $N=1$. In that
case the action is
\ba
S = -\frac{1}{8\pi^2} \int  \!dt \,dx \left( \phantom{{x^2}^2}\!\!\!\!\!\!\!\!\!\kappa
\,\phi\,\partial_x(\partial_t-v\partial_x)\phi \,+
\right.
\n+
\left.
\kappa'
\phi\,(\partial_t-v\partial_x)(\partial_x\phi)^2 \phantom{{x^2}^2}\!\!\!\!\!\!\!\!\!\right)
\label{action}
\ea
where we have dropped the subindexes since we have a single field.
This is a chiral boson theory with a cubic higher derivative self
interaction, the latter arising essentially from the smoothness of
the edge.

To see the effect of this last term in the physical properties, we
compute the propagator since it is directly related to the
tunnelling exponent \cite{Wen}.

In Fourier space we write for the Feynmann propagator
\be G(p,\omega_p) = \frac {2\pi} {\kappa
p\,(\omega_p-vp)+i\epsilon} \label{propa} \ee
while the vertex proportional to
\ba -\kappa'\delta(p+q+r)(\omega_p-vp)\,q\,r \ea
Then, the one loop corrected propagator reads
\ba
G'(p,\omega_p) &=&\frac {2\pi} {\kappa p\,(\omega_p-vp)-\delta
G(p,\omega_p)}
\ea
We consider those Feynmann diagrams whose contribution to the
corrected propagator is
\ba
\delta G(p,\omega_p) =\pi\Lambda^2\frac{\kappa'^2}{\kappa^2 }
(\omega_p-v p) p
\ea
where $\Lambda$ is a momentum cutoff. Replaced in (\ref{propa})
this gives

\be
G(p,\omega_p) = \frac 1 {(\kappa +\delta
\kappa)p\,(\omega_p-v p)}
\ee
where the non-universal correction to the level is given by

\ba
\delta\kappa &=& -\pi \frac{\kappa'^2}{\kappa^2}\Lambda^2
\ea
A non vanishing correction to the velocity arises from the
remaining diagrams but it produces no physical consequences.

Since the tunnelling exponent is directly related to the level by
$\alpha\approx\kappa$, the appearance of a correction for the
latter implies, to first order, a correction to the tunnelling
exponent. This is our main result.

To estimate the magnitude of this correction, we need to relate
the momentum cutoff $\Lambda$ to the minimal space distance
measurable $a$, which naturally leads to the identification
$\Lambda = \pi/a$. Using (\ref{levels}) we then get for this
choice of the cutoff

\be \delta\kappa =  - \frac{\pi}{(4a^2\rho_0)^2} = -\frac{\pi}{16}
\approx -0.196 \ee
where we have further identified $\rho_0=1/a^2$. The predicted
dependence on the density of this non-universal correction could
in principle be tested experimentally.

Putting all together this leads to a linear dependance of the
tunnelling exponent $\alpha$ as a function of $1/\nu$
%\be \alpha(\frac{1}{\nu}) = \frac{1}{\nu} -\frac{\pi}{16} \ee
%
%$\alpha \approx
%\frac{1.16}{\nu} -0.58$
which is very close to the experimental fit presented in
\cite{first diff} (see for example Fig.\ 3 of this reference). It
can be observed that the departure of our result from the
experimental data becomes greater for increasing $1/\nu$, which
could be attributed to the need to work with $N>1$ in (\ref{cb}).
This could in turn be related to the need to consider a wider
boundary region $W > a$. The analysis of the higher $N$ case will
be presented separately \cite{CG}.

To summarize, we have constructed a boundary effective field
theory for a fractional quantum Hall effect droplet by including
the granularity of the fluid (the interparticle distance $a$) and
the effects of a smooth boundary, starting from the microscopic
model. By further assuming a space dependent density to define the
boundary region, we have computed the correction to one loop order
of the tunnelling exponent which is good agreement with
experimental results for the case of an almost sharp boundary.
%This represents an improvement with respect to previous analytical
%approaches, since the correction obtained provides a result
%closer to experimental values

Wider boundary regions can be easily treated within the present
approach, by simply taking higher values of $N$ in (\ref{cb}).

Other approaches to include a boundary in a Chern-Simons
description of a quantum Hall droplet have been proposed
\cite{Poly}. In this context, it would be interesting to study the
connection between the two approaches.

It should be mentioned that the experimental data in \cite{first
diff} was reanalyzed in \cite{plateau} in view of the results of
\cite{Chamon-Fradkin}, and certain degree of agreement with the
plateau structure was obtained. Here we propose an alternative
description which provides a reasonable agreement with the raw
experimental data as presented in \cite{first diff}. It would be
interesting to study the higher order theory that we have
constructed in the present Letter along the lines of the approach
presented in ref.\ \cite{Chamon-Fradkin}. This could provide a
closer agreement between theoretical and experimental results.

It would be also interesting to analyze the consequences of the
higher order corrections induced by the smoothness of the boundary
that we have obtained here in the transition between plateaux
states.

%NEG cita a la version larga (como la mia)
%
The approach proposed in this note has been formalized carefully starting 
from a microscopic description and developped further in \cite{larg}.

\acknowledgments We thank A.\ Dobry, E.\ Fradkin, J.\ Polonyi, P.\
Pujol and G.\ Silva for helpful discussions. N.E.G. wants to thank
the Laboratoire de Physique Th\'{e}orique of the Universit\'{e}
Louis Pasteur de Strasbourg for hospitality and financial support
during the early stages of this work. This work was partially
supported by ECOS-Sud Argentina-France collaboration (Grant
A04E03) and PICS CNRS-CONICET (Grant 18294).

\end{document}